\documentstyle[14pt,epsf,rotate]{article}
\newcommand{\Tr}{\mathop{\rm Tr}\nolimits}
\newcommand{\Arg}{\mathop{\rm Arg}\nolimits}

\begin{document}
\author{A.~A. Slavnov \thanks{e-mail:$~~$ slavnov@mi.ras.ru} \ \ and
\ \ N.~V. Zverev \thanks{e-mail:$~~$ zverev@mi.ras.ru} \\
{\it Steklov Mathematical Institute, Russian} \\
{\it Academy of Sciences, Vavilov st. 42, 117966, Moscow, Russia} \\
{\it Moscow State University, Physical Faculty,} \\
{\it 117234, Moscow, Russia}
}
\title{
Fermion theories on a 2d torus with Wilson action
improved by Pauli -- Villars regularization
}
\date{}
\maketitle

\begin{abstract}
Vectorial and anomaly free chiral U(1) fermion models on a 2d finite
lattice are considered. It is demonstrated both numerically and
analytically that introduction of Pauli -- Villars type regularization
supresses the symmetry breaking effects caused by the Wilson term.
\end{abstract}

\newpage

\section {Introduction}

Treating chiral fermions on the lattice still presents serious problem.
Due to well known "no-go" theorem \cite{NiNe} there is no straightforward
way to remove fermion spectrum degeneracy without breaking chiral invariance.
Several possibilities to overcome this difficulty were proposed (for recent
review see \cite{Sh}), but all of them have certain problems. A really
successful model has to provide reasonable results both in perturbative
and nonperturbative regions at least for sufficiently small lattice spacing.

Existing computer facilities allow efficient nonperturbative tests only in
two-dimensional models, the U(1) gauge model on the torus being a popular
example.

In the  present paper we apply to the toron model the method, proposed in
ref.\cite{FSl}. The idea of this method is to introduse in the lattice models
additional gauge invariant Pauli -- Villars (PV) type regularization which
supresses the contribution of momenta close to the border of Brillouin zone,
$|p| \!\sim\! \pi / a$. Pauli -- Villars fields mass $M$ introduces a new
scale which is choosen to satisfy the condition $M \!\ll\! a^{-1}$. With this
condition fulfilled any modification of the action at the distances of order
$a$ becomes irrelevant. Therefore, if the additional regularization respects
vectorial and chiral gauge invariance, possible symmetry breaking effects due
to introduction of Wilson term or some other device removing spectrum
degeneracy, are supressed and vanish in the continuum limit.

In the paper \cite{FSl} it was shown that introducing PV type regularization
together with the standard Wilson term \cite{W} for anomaly free chiral gauge
models on the infinite lattice one gets in the framework of perturbation
theory correct continuum results without chiral noninvariant counterterms.
Possible chirality breaking effects are of order $a$.

Below we shall check these results nonperturbatively for the two dimensional
model -- U(1) gauge invariant interaction of fermions on the 2d finite
lattice. To compare our calculation with known exact results for continuum
theory on the torus [5 -- 7] the gauge field is
chosen to be constant. We present the results both for vectorial and
anomaly free chiral models and compare them with the corresponding
calculations for Wilson fermions supplemented by gauge noninvariant
counterterms as proposed by the Roma group \cite{Roma}.

\section {Vectorial lattice model}

We start with the vectorial model on the finite lattice described by the
action \cite{W}
\begin{equation}\label{1}
I_{VW} = \frac{1}{2} \sum_{x, \mu} \left\{ \overline{\psi}(x)\gamma_\mu U_\mu
\psi(x+\hat{\mu}) - \overline{\psi}(x) \left[ \psi(x+\hat\mu)-\psi(x)
\right] \right\} + {\rm h.c.} \end{equation}
Here $-N/2+1 \!\le\! x_\mu \!\le\! N/2$, $\mu$=0, 1; $N$ is the number of
lattice sites. The lattice spacing is chosen to be equal to 1. The first term
describes gauge invariant interaction of fermions with the constant field
$$
U_\mu=\exp\left( \frac{2\pi{\rm i}}{N}h_\mu \right).
$$
The second term is the Wilson term removing fermion spectrum degeneracy.
Being interested in the study of symmetry breaking effects on a lattice we
choose the gauge noninvariant Wilson term. Note that chiral invariance is
broken both for noninvariant or covariant Wilson terms. The Fermi field
$\psi$ satisfy antiperiodic boundary conditions.

A straightforward calculation gives the following expression for the
determinant:
\begin{equation}\label{2}
D_{VW} = \prod_{p=-N/2+1}^{N/2} \frac{B^2(p, h)+W^2(p)}{B^2(p, 0)+W^2(p)},
\end{equation}
where
\begin{equation}\label{3}
B_\mu(p, h) = \sin \frac{2\pi}{N}\left( p_\mu-h_\mu-1/2 \right),
\end{equation}
\begin{equation}\label{4}
W(p) = \sum_{\mu=0}^{1} \left( 1-\cos\frac{2\pi}{N} \left( p_\mu-1/2
\right) \right).
\end{equation}
The determinant (\ref{2}) is normalized to 1 at $h$=0. The corresponding
expression in the continuum theory looks as follows [5 -- 7]:
\begin{equation}\label{5}
D_{VC} = {\rm e}^{-2\pi h_1^2}\prod_{n=1}^{\infty} \Bigl| F[n, h] F[n, -h]
\Bigr|^2.
\end{equation}
Here
$$
F[n, h]=\frac{{\displaystyle 1 + {\rm e}^{-2\pi(n-1/2)+2\pi{\rm i}
(h_0+{\rm i}h_1)}}} {{\displaystyle 1 + {{\rm e}^{-2\pi(n-1/2)}}}}.
$$

The determinants (\ref{2}), (\ref{5}) satisfy the following
symmetry properties:
$$
D[h_0, h_1] = D[h_1, h_0] = D[-h_0, h_1] = D[h_0, -h_1].
$$
It allows to consider the fields $h_\mu$ in the interval $0 \!\le\! h_0 \!
\le\! h_1$. The continuum determinant (\ref{5}) satisfies also periodicity
condition
$$
D_{VC}[h_0, h_1] = D_{VC}[h_0+n_0, h_1+n_1], \qquad n_0, n_1=0, \pm 1,
\pm 2, \dots
$$
Due to the breaking of gauge invariance by the Wilson term, the lattice
determinant satisfies weaker periodicity condition
$$
D_{VW}[h_0, h_1] = D_{VW}[h_0+\frac{N}{2}n_0, h_1+\frac{N}{2}n_1].
$$

Computer simulations were performed for $0 \!\le\! h_0 \!\le\! 0.5$,
$-0.5 \!\le\! h_1 \!\le\! 1.5$ using the lattices with $N$=32, 160. The
results are presented at Fig.1 (curves 1, 2). One sees that the lattice
results do not agree with the continuum ones in the whole interval of $h_0$,
$h_1$. It shows that although formally the action (\ref{1}) has the correct
continuum limit, symmetry breaking effects due to noninvariant Wilson term
change quantum corrections drastically.

\begin{figure}[tb]
\epsfxsize=16cm

\vspace*{-7mm}
\hspace*{-3mm}
\rotate[r]{\epsfbox{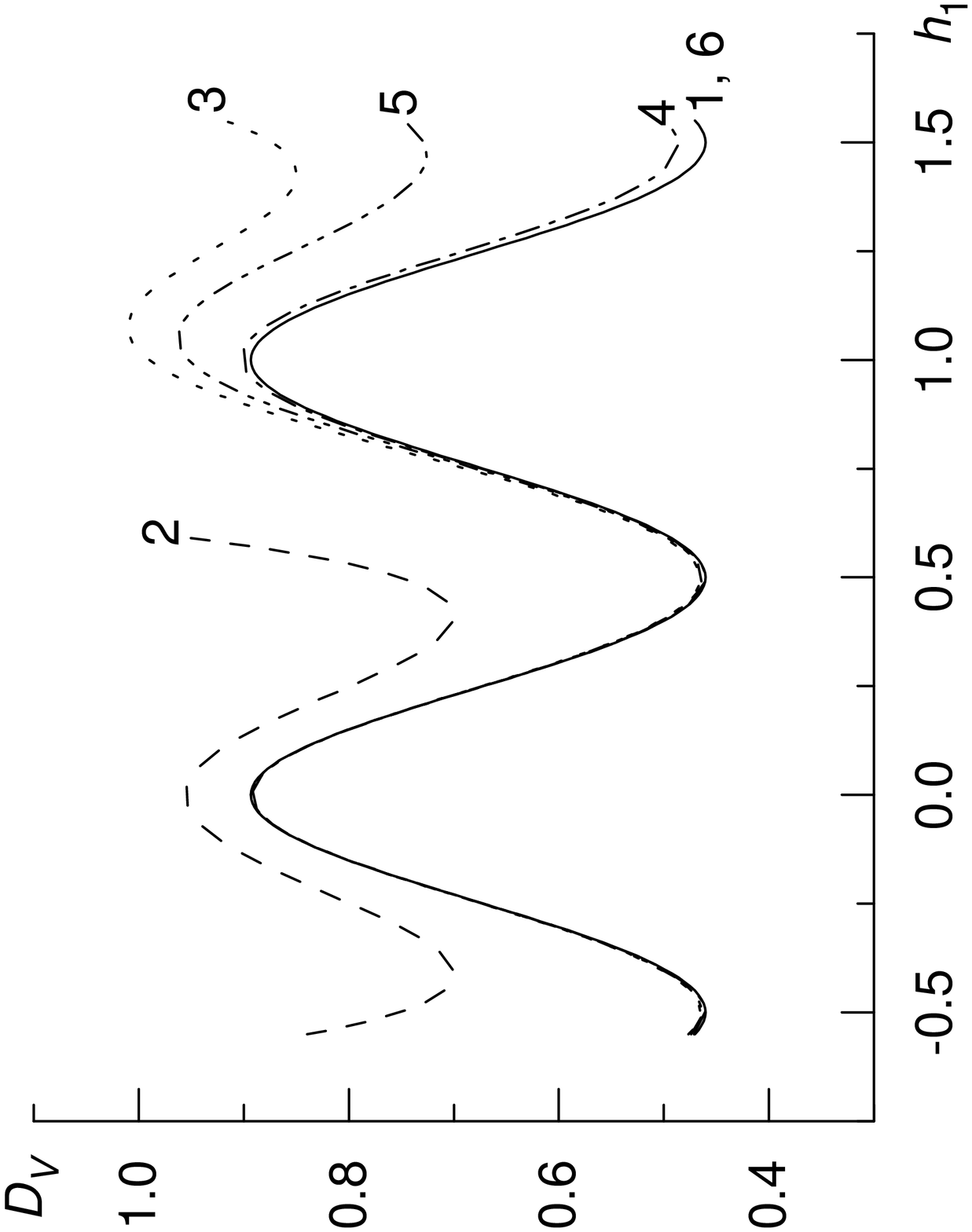}}

\vspace*{-48mm}
{\bf Fig.1.}
Vectorial determinants $D_V$ as functions of $h_1$ at $h_0=0.2$: \\
\hspace*{16mm} 1 -- $D_{VC}$ on the torus; \quad  2 -- $D_{VW}$ without
PV field, $N=160$; \\
\hspace*{16mm} 3, 4 -- $D_{VR}$ with PV field, $M=M_2$: \quad 3 -- $N=32$,
\quad  4 -- 160; \\
\hspace*{16mm} 5, 6 -- $D_{VK}$ with counterterm: \quad   5 -- $N=32$,
\quad 6 -- 160
\bigskip
\end{figure}

Below we show that this discrepancy vanishes if one modifies the action
(\ref{1}) by introducing additional PV regularization. The modified action
looks as follows:
$$
I_{VR} = I_{VW}+ I_{PV},
$$
\begin{eqnarray*}
I_{PV} = \frac{1}{2} \sum_{x} \biggl\{ \sum_{\mu}
\left[ \overline{\phi}(x)\gamma_\mu U_\mu
\phi(x+\hat{\mu}) - \overline{\phi}(x) \Bigl( \phi(x+\hat{\mu})-\phi(x)
\Bigr) \right] + \biggr.\\
\biggl. + M\bar{\phi}(x)\phi(x)\biggr\} + {\rm h.c.}
\end{eqnarray*}
Here $\phi(x)$ is a bosonic PV field having the same spinorial and internal
structure as $\psi$. In the considered model one PV field is sufficient to
supress the contribution of the region near the border of the Brillouin zone.

Regularized determinant may be presented in the form
$$
D_{VR} = D_{VW}[h]D_{PV}[h],
$$
where $D_{VW}[h]$ is given by eq.(\ref{2}) and $D_{PV}$ is the corresponding
expression for the PV field:
\begin{equation}\label{5a}
D_{PV}[h] = \prod_{p=-N/2+1}^{N/2} \left( \frac{B^2(p, h)+\left(W(p)+M
\right)^2}{B^2(p, 0)+\left(W(p)+M\right)^2} \right)^{-1}.
\end{equation}

\begin{figure}[tb]
\epsfxsize=16cm

\vspace*{-7mm}
\hspace*{-3mm}
\rotate[r]{\epsfbox{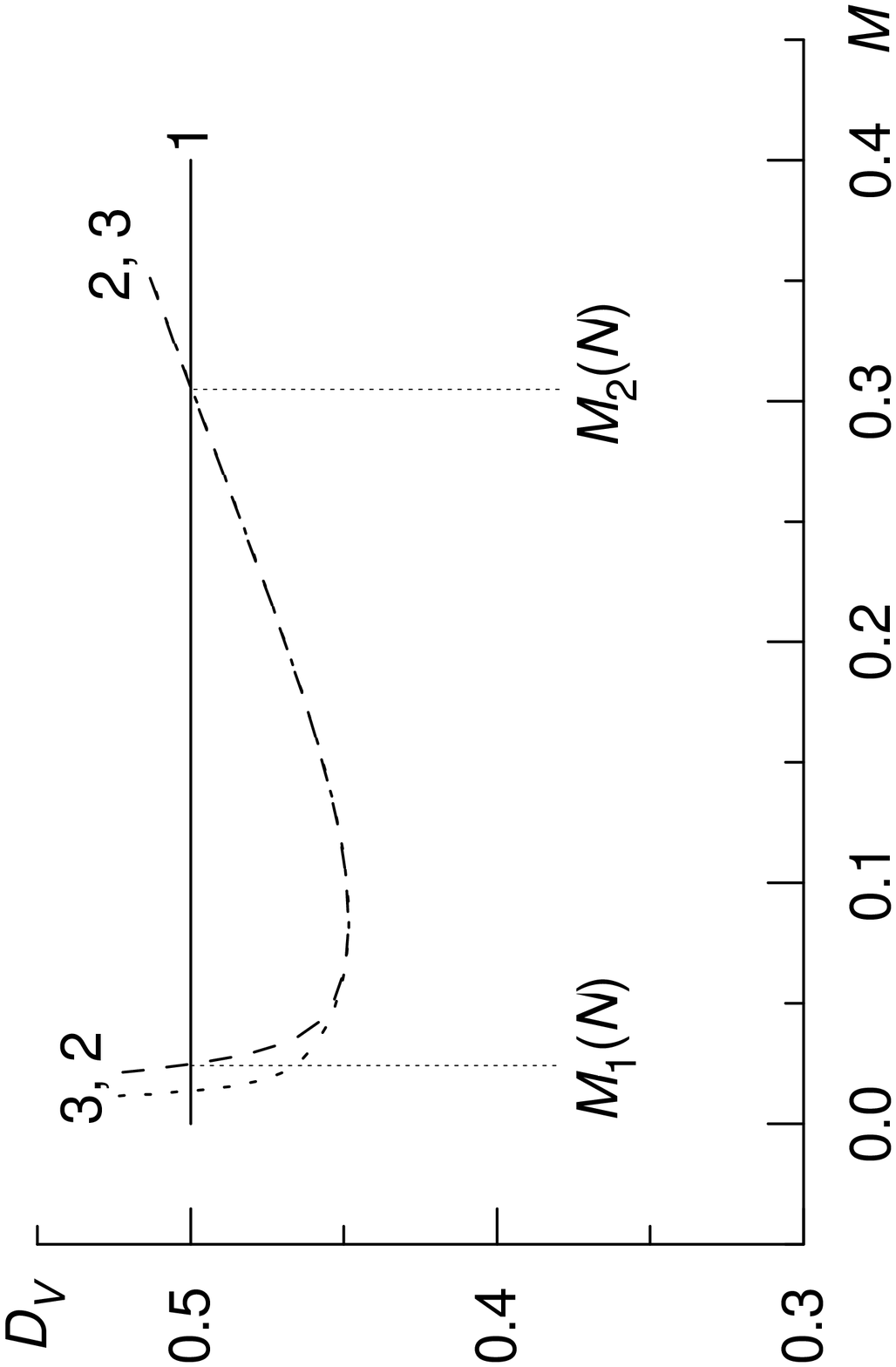}}

\vspace*{-63mm}
{\bf Fig.2.}
Vectorial determinants $D_V$ as functions of $M$ \\
\hspace*{16mm} at $h_0=0.2$ and $h_1=0.4$: \\
\hspace*{16mm} 1 -- $D_{VC}$ on the torus; \quad  2, 3 -- $D_{VR}$ with PV
field: \\
\hspace*{16mm} 2 -- $N=160$, \qquad  3 -- 320
\bigskip
\end{figure}

Note that in this equation we defined the determinant of a bosonic PV field
as inverse of the corresponding fermion determinant. One can wonder if this
determinant may be really presented as a path integral of the corresponding
bosonic action $I_{PV}$. A naive integral over bosonic fields of
$\exp(-I_{PV})$ does not exist as the action is not positive definite.
However as eq.(\ref{2}) shows the fermionic determinant $D_{VW}$ is positive
and can be presented as a path integral of an exponent of a positive
definite action. Therefore the corresponding PV determinant also can be
written as a path integral of an exponent of a positive action. Its explisit
form can be easily read of the eq.(\ref{5a}). In general, if the modulus of
a fermion determinant can be presented as a path integral of exponent of a
positive local action, analogous representation is valid for the
corresponding PV determinant.

The results of calculations for regularized determinant as function of the
regularizing mass $M$ for fixed $h_0$, $h_1$ at $N$=160 and 320 are
presented at Fig.2. Exact agreement with the continuum is achieved for two
values of $M$=$M_1$, $M_2$. However in the whole interval
$M_1 \!<\! M \!<\! M_2$ which in our case is of order $M_1$=0.01 -- 0.03 and
$M_2 \!\approx\! 0.3$ the discrepancy is within $10\%$. So to get the correct
result one needs to tune the regularizing mass, but it is not really a fine
tuning.

The dependence of the regularized determinant on the $h_1$ at $h_0$=0.2
for value $M$=$M_2$ are shown at Fig.1 (curves 3, 4). One sees that for
$-0.5 \!\le\! h_\mu \!\le\! 0.5$ there is a very good agreement with the
continuum results both for $N$=32 and 160. To get a good agreement for larger
$h$ one needs a bigger lattice: for $0.5 \!\le\! h_1 \!\le\! 1.5$ $N$=160
provides a good agreement whereas $N$=32 gives a sizable mistake.

The reason for that is easily understood. PV regularization supresses the
contribution of momenta for which $B^2+W^2 \!\gg\! M^2$. One sees from the
eq.(\ref{3}) that $B_\mu$ depends only on the difference $p_\mu - h_\mu$.
Hence for $h$ big enouth $B^2$ can be much bigger than $M^2$ even for
relatively small $p$. In this way part of a physical region will be also
supressed. This effect obviously disappears with increasing of $N$. It seems
to be a common problem for any non gauge invariant lattice scheme. In
particular our calculations show that the same phenomenon occurs in Roma
approach \cite{Roma}. In the particular model one can get better agreement
by using other approaches e.g. overlap formalism \cite{NNe} or staggered
fermions as was done in somewhat different context in ref.\cite{BHS}. In our
case we could easily avoid the problem by using a gauge invariant Wilson
term. In this case our regularized action is manifestly gauge invariant and
the determinant is periodic with the period 1. Hence if there is an agreement
with the continuum in the interval $-1/2 \!\le\! h_\mu \!\le\! 1/2$ it is
continued automatically to any $h$.

However for general chiral models considering large nonsmooth external fields
really makes a problem. One way to avoid it is to supress the contribution of
such fields by some additional regularization. In the paper \cite{FSl} it was
shown that in perturbation theory higher covariant derivative for gauge field
may play this role. It would be important to check if it also in
nonperturbative calculations.

\begin{figure}[tb]
\epsfxsize=16cm

\vspace*{-7mm}
\hspace*{-3mm}
\rotate[r]{\epsfbox{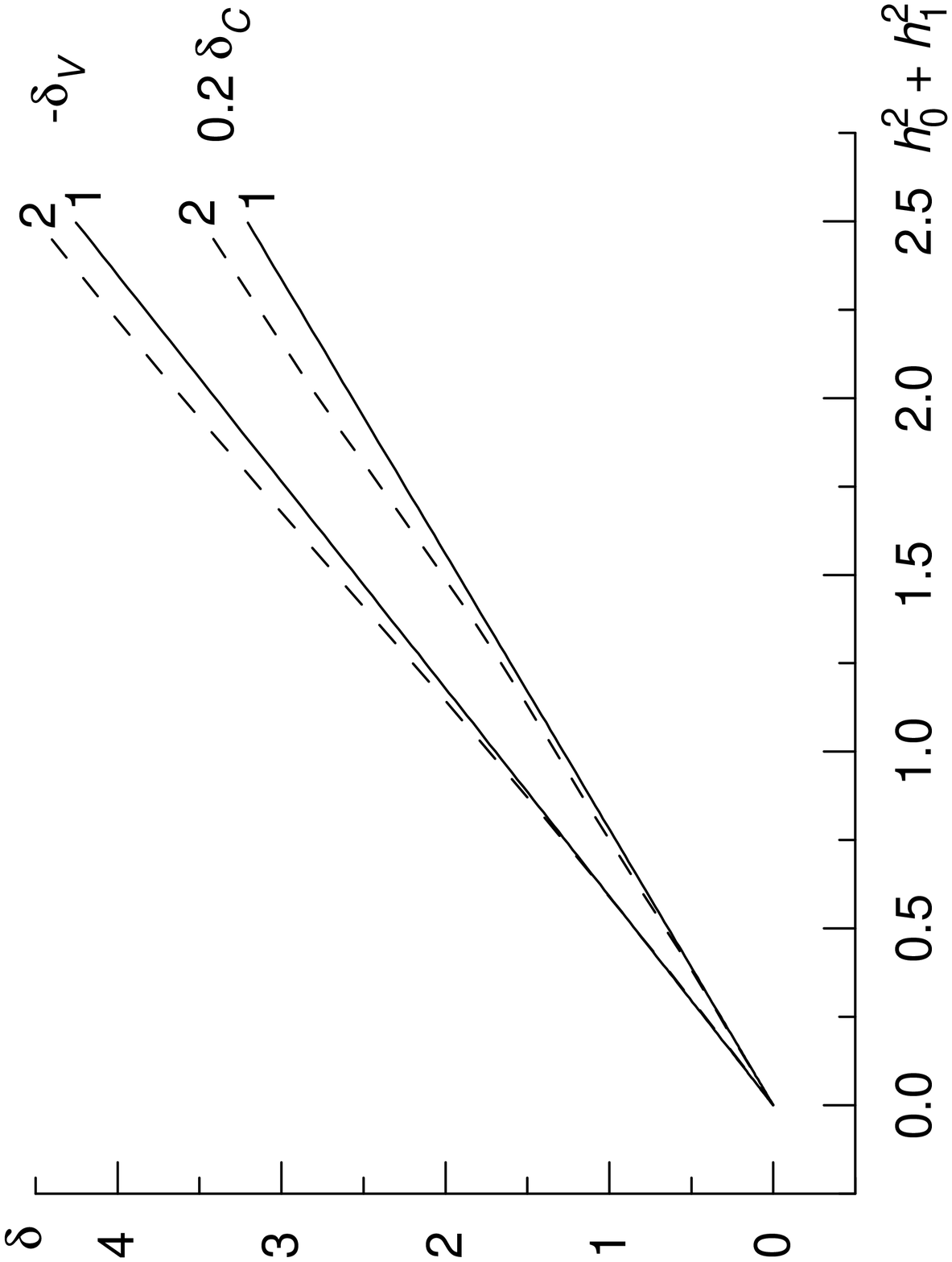}}

\vspace*{-48mm}
{\bf Fig.3.}
Values $- \delta_{V}$ and $0.2 \delta_{C}$ as functions
of $h_0^2 + h_1^2$: \\
\hspace*{16mm} 1 -- $N=160$, $h_0=0$ and 0.2; \qquad  2 -- 32, 0
\bigskip
\end{figure}

The alternative way to improve the agreement with the continuum results is,
following the Roma approach \cite{Roma}, to introduce to the action (\ref{1})
a gauge noninvariant counterterm. Introducing the counterterm
$k_V\left(h_0^2+h_1^2\right)$, one gets for the determinant
\begin{equation}\label{6}
D_{VK} = D_{VW}\exp\left[k_V\left(h_0^2+h_1^2\right)\right].
\end{equation}
To determine the coefficient $k_V$ we calculated the value $\delta_V \!=\!
-\ln (D_{VW} / D_{VC})$ as a function of $h_0^2+h_1^2$, the result being
presented at Fig.3. It gives $k_V \!=\! -1.6877$. Substituting this value to
eq.(\ref{6}) one gets for $D_{VK}$ the curves presented at Fig.1 (curves
5, 6).  Qualitatively we have the same situation as in the case of PV
regularization:  for $N$=32 there is a good agreement in the interval $-0.5
\!\le\! h_\mu \!\le\! 0.5$, for $N=160$ the interval of agreement extends
to $-0.5 \!\le\! h_\mu \!\le\! 1.5$.

Therefore in the model under consideration both PV regularization and Roma
approach give analogous results. The advantages of PV approach are twofold.
First of all in this case the only source of gauge symmetry breaking of the
action is the Wilson term and no gauge noninvariant counterterms are needed.
One can avoid gauge symmetry breaking completely by choosing other methods
of spectrum degeneracy removing. SLAC-model supplemented by PV regularization
is the example of manifestly gauge invariant (although nonlocal) model without
spectrum doubling \cite{Sl}. Moreover, gauge symmetry preserving PV type
regularization may be constructed also for anomaly free chiral models
\cite{FSl, Sl, FrSl}.

Secondly, in more complicated four dimensional models in the framework of
Roma approach one would need to introduce more gauge noninvariant
counterterms, whereas PV type regularization cuts all unwanted effects of
momenta $p \!\sim\! \pi /a$, provided the Yang -- Mills fields are also
regularized by means of higher covariant derivatives (see the discussion in
\cite{FSl}). It is worthwhile to mention that even in a two-dimensional
models a simple mass renormalization may be not sufficient to get a good
agreement with the continuum \cite{BHS}.

The numerical results obtained above are in a good agreement with the
analitical estimaties. One can show that all the lattice diagrams with more
than two external lines differ from the corresponding diagrams in the
continuum toron model by the terms of order ${\rm O}(1/MN)$. At the same time
$$
\Pi_{VW}(0) = \Pi_{VC}(0) - \frac{16\pi^2}{27\sqrt{3}},
$$
where $\Pi_{VC}(0)\!=\!2\pi$ is the mass gap in the comtinuum toron model.
Therefore to get the correct continuum result without PV regularization
one has to add to determinant $D_{VW}$ the counterterm with the coefficient
$k_V \!=\! -8\pi^2 / 27\sqrt{3} \!\approx\! -1.6884$, to be compared with the
numerical value $-1.6877$.

If the PV fields are introduced the following expressions for polarization
operator can be derived:
$$
\Pi_{VR}(0) = 2\pi + {\rm O}(1/MN) + {\rm O}(M \ln^2 N), \qquad
\mbox{if } M \to 0 \mbox{ when } N \to \infty;
$$
$$
\Pi_{VR}(0) \to 2\pi, \qquad \mbox{if } M = 0.307 \mbox{ when }
N \to \infty.
$$

These values are in a good agreement with the numerical results presented at
Fig.2. Analogous estimates are done for anomaly free chiral models.

\section {11112 lattice model}

In this section we apply the approach described above to anomaly free
chiral models on the finite lattice. The possibility to supress the
contribution of momenta close to the border of Brillouin zone in anomaly
free 4d models by introducing a chiral gauge invariant PV type regularization
is related to the fact that in this case divergent diagrams (with less than 5
external vector lines) contribute only to the modulus of the determinant.
From the point of view of these diagrams the theory is essentially vector
like. In two dimension the anomalous diagram is the polarization operator,
and the only way to get rid off anomaly is to use a combination of left-handed
and right-handed interactions in which the terms proportional to $\gamma_5$
cancel. In particular one can consider 11112 model including four left-handed
fermions with charge 1 and one right-handed fermion with charge 2.

The action of this model can be written in the form \cite{W}
\begin{eqnarray}\label{7}
I_{CW} = \frac{1}{2}\sum_{k=1}^{4}\biggl\{ \Bigl[ \sum_{x, \mu}
\overline{\psi}_{k+}(x)\gamma_\mu\left(P_+ U_\mu + P_- \right)\psi_{k+}
(x+\hat{\mu}) - \overline{\psi}_{k+}(x) \times \Bigr. \biggr. \nonumber \\
\biggl. \Bigl. \times\Bigl( \psi_{k+}(x+\hat{\mu}) - \psi_{k+}(x) \Bigr)
\Bigr] \biggr\} + \frac{1}{2}\sum_{x, \mu}\Bigl[ \overline{\psi}_{-}(x)
\gamma_\mu (P_{+} + U_\mu^2 P_{-}) \Bigr.\times \nonumber \\
\times \Bigl.\psi_{-}(x+\hat{\mu}) +\overline{\psi}_{-}(x)
\Bigl( \psi_{-}(x+\hat{\mu})- \psi_{-}(x)\Bigr) \Bigr] + {\rm h.c.}
\end{eqnarray}
Here $P_{\pm} \!=\! \frac{1}{2}(1 \pm \gamma_3)$.
To supress the contribution of the region near the border of the Brillouin
zone we introduce the following interaction of PV fields:
\begin{eqnarray}\label{8}
I_{PV}=\frac{1}{2}\sum_{x, \mu}\Bigl\{ \overline{\Phi}(x)
\gamma_\mu\left( P_R + U_{\mu}^2 P_L \right) \Phi(x+\hat\mu)
- \overline{\Phi}(x)\times\ \Bigr. \nonumber \\
\Bigl. \times \left[ \Phi(x+\hat\mu) - \Phi(x) \right] \Bigr\}
+\frac{1}{2}\sum_{x}M\overline{\Phi}(x)\tau_1 \Phi(x) + {\rm h.c.}
\end{eqnarray}
Here $\Phi(x)$ is doublet of bosonic PV fields $\varphi_+$, $\varphi_-$
having the same spinorial structure as $\psi_+$, $\psi_-$ and satisfying
antiperiodic boundary condition; $P_L \!=\! \frac{1}{2}(1+\gamma_3 \tau_3)$,
$P_R \!=\! \frac{1}{2}(1-\gamma_3 \tau_3)$; $\tau_1$, $\tau_3$ are Pauli
matricies. The regularized action
$$
I_{CR}=I_{CW}+I_{PV}
$$
in the formal continuum limit, when the Wilson terms are neglected, is
invariant under chiral gauge transformations of the fields $\psi_\pm$,
$\Phi$.

The action $I_{CR}$ generates the following propagators:
\begin{equation}\label{9}
S_{\psi_{+} \overline{\psi}_{+}} = S_{\psi_{-} \overline{\psi}_{-}} =
\frac{{\rm i}\hat{B}(p)+W(p)}{B^2(p)+W^2(p)},
\end{equation}
\begin{equation}\label{10}
S_{\Phi \overline{\Phi}} = \frac{{\rm i}\hat{B}(p)+W(p)+M}{B^2(p)+\left(W(p)
+M\right)^2}\cdot\frac{1+\tau_1}{2} + \frac{{\rm i}\hat{B}(p)+W(p)-M}{B^2(p)
+\left(W(p)-M\right)^2}\cdot\frac{1-\tau_1}{2},
\end{equation}
where $\hat{B}(p) \!=\! \sum\limits_{\mu=0}^{1}B_\mu(p)\gamma_\mu$, $B_\mu(p)
\!=\! B_\mu(p, 0)$ is given by eq.(\ref{3}) and $W(p)$ is given by (\ref{4}).

The polarization operator is a sum of the terms which have a form
$$
\Pi_{\mu\nu} \sim \sum_{p=-N/2+1}^{N/2} q^2 \Tr{ \left(\gamma_\mu P_\pm
V_\mu (p) S(p) \gamma_\nu P_\pm V_\nu (p) S(p)\right)},
$$
where $V_\mu (p)$ is a lattice vertex function and the propagators $S$ are
given by eqs.(\ref{9}), (\ref{10}).

Consider firstly the contribution of physical fields. As the model is
anomaly free, the terms proportional to $\gamma_3$ cansel and one gets
$$
\Pi_{\mu\nu}^{CW} \sim \sum_{p=-N/2+1}^{N/2} 4 \Tr{ \left(\gamma_\mu
V_\mu (p) \frac{{\rm i}\hat{B}(p)+W(p)}{B^2(p)+W^2(p)} S(p) \gamma_\nu
V_\nu (p) \frac{{\rm i}\hat{B}(p) +W(p)}{B^2(p)+W^2(p)}S(p)\right)}.
$$

Following the reasonings of ref.\cite{FSl} one can show that analogous
expression for PV fields cancel the contribution of momenta close to the
border of the Brillouin zone and the remaining part in the limit $N \!\to\!
\infty$ coincides with the gauge invariant continuum result.
Separating the summation domain into two parts $V_{{\rm in}}$:  $|p| \!<\!
N^\gamma$, $\gamma \!<\! \frac{1}{2}$, and $V_{{\rm out}}$:  $|p| \!\ge\!
N^\gamma$, and choosing $M \!\sim\! N^\delta$, $\delta \!<\! \gamma$, one
sees that in the domain $V_{{\rm out}}$ one can expand the PV fields
propagators in terms of $M^2$. The zero order term coincides with the
propagator of $\psi$-fields and due to our choice of charges the leading
terms in the expansion of $\Pi_{\mu\nu}^{PV}$ cancel the value
$\Pi_{\mu\nu}^{CW}$. The next term is majorated by $N^{-2\gamma}M^2$ and
vanishes in the limit $N \!\to\! \infty$.

In the domain $V_{{\rm in}}$ one can expand the Wilson term in power of $p$.
The leading term in the limit $N \!\to\! \infty$ gives the gauge invariant
continuum expression and the higher order terms are majorated by
$N^{4\gamma -2} \!\to\! 0$ when $N \!\to\! \infty$.

By the same reasoning all higher order diagrams, which correspond to power
counting convergent integrals in the limit $N \!\to\! \infty$ coincide with
the corresponding continuum expressions. So we expect that simulations of the
11112 model (\ref{7}) regularized by the PV action (\ref{8}) has to provide a
reasonable approximation to the continuum toron model.

A straightforward calculation gives for the regularized lattice 11112
determinant normalized on 1 at $h$=0 the following expression:
\begin{equation}\label{11}
D_{CR} = D_{CW}[h] D_{PV}[h],
\end{equation}
$$
D_{CW} = D_{+W}^4[h] D_{+W}^*[2h],
$$
$$
D_{+W}[h] = \prod_{p=-N/2+1}^{N/2} \frac{G[p, h]}{H[p, 0]},
$$
$$
D_{PV}[h] = \prod_{p=-N/2+1}^{N/2} \left( \frac{\bigl| G[p, 2h] \bigr|^2 +
M^2\bigl[ B^2(p, 2h) + B^2(p, 0) - 2W^2(p) + M^2 \bigr]} {H[p, M]
H[p, -M]} \right)^{-1}.
$$
Here
$$
G[p, h] = [B_0(p, h) + {\rm i}B_1(p, h)] [B_0(p, 0)
- {\rm i}B_1(p, 0)] + W^2(p),
$$
$$
H[p, {\cal M}] = B^2(p, 0) + \bigl( W(p) + {\cal M} \bigr)^2,
$$
$B_\mu(p, h)$, $B^2(p, h)$ and $W(p)$ are defined above. One sees that the
regularization changes only modulus of the lattice determinant whereas its
argument remains intact, i.e. $\Arg D_{CR} \!=\! \Arg D_{CW}$.

The corresponding expression in the continuum toron 11112 theory looks as
follows [5 -- 7]:
\begin{equation}\label{12}
D_{CC}[h] = D_{+C}^4[h] D_{+C}^*[2h].
\end{equation}
Here
$$
D_{+C}[h] = {\rm e}^{{\rm i}\pi h_1 (h_0 + {\rm i}h_1)}
\prod_{n=1}^{\infty} F[n, h] F[n, -h],
$$
and expression for the $F[n, h]$ is given above.

\begin{figure}[tb]
\epsfxsize=16cm

\vspace*{-7mm}
\hspace*{-3mm}
\rotate[r]{\epsfbox{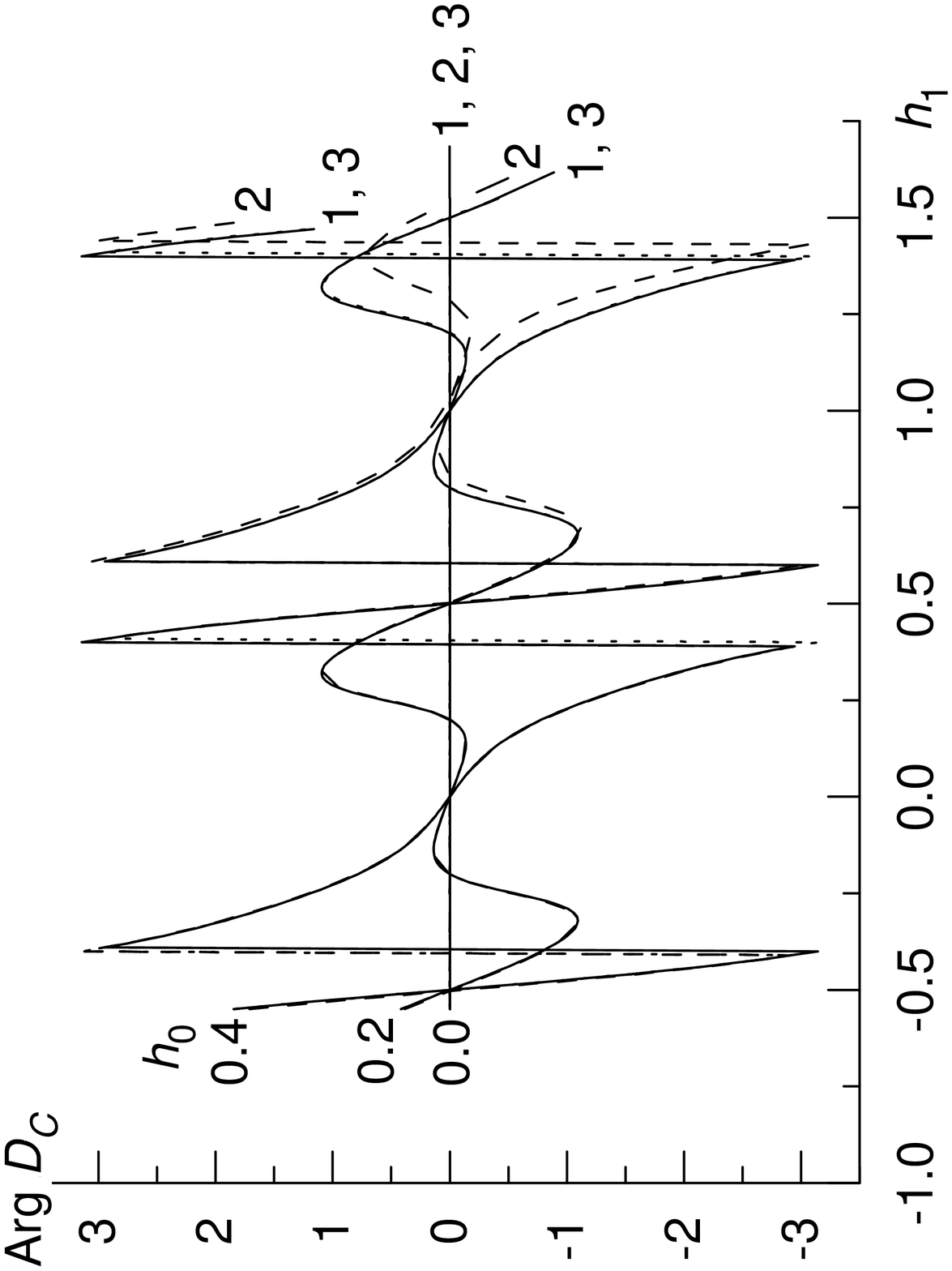}}

\vspace*{-48mm}
{\bf Fig.4.}
11112 determinant arguments ${\rm Arg}\,D_C\,$ as functions \\
\hspace*{16mm} of $h_1$ at different $h_0$: \\
\hspace*{16mm} 1 -- ${\rm Arg}\,D_{CC}\,$ on the torus; \\
\hspace*{16mm} 2, 3 -- $\Arg D_{CR}$=$\Arg D_{CK}$=$\Arg D_{CW}$:
\quad 2 -- $N=32$, \quad  3 -- 160

\bigskip
\end{figure}

The determinants (\ref{11}), (\ref{12}) satisfy the following symmetry
properties:
$$
D[h_0, h_1] = D^*[h_1, h_0] = D^*[-h_0, h_1] = D^*[h_0, -h_1],
$$
$$
D_{CC}[h_0, h_1] = D_{CC}[h_0+n_0, h_1+n_1], \qquad n_0, n_1 =0, \pm 1,
\pm 2, \dots
$$
\begin{equation}\label{13}
D_{CR}[h_0, h_1] = D_{CR}[h_0 + N n_0, h_1 + N n_1].
\end{equation}
Due to these properties it is sufficient to take the external fields in the
interval $0 \le h_0 \!\le\! h_1 \!\le\! N/2$.

\begin{figure}[tb]
\epsfxsize=16cm

\vspace*{-7mm}
\hspace*{-3mm}
\rotate[r]{\epsfbox{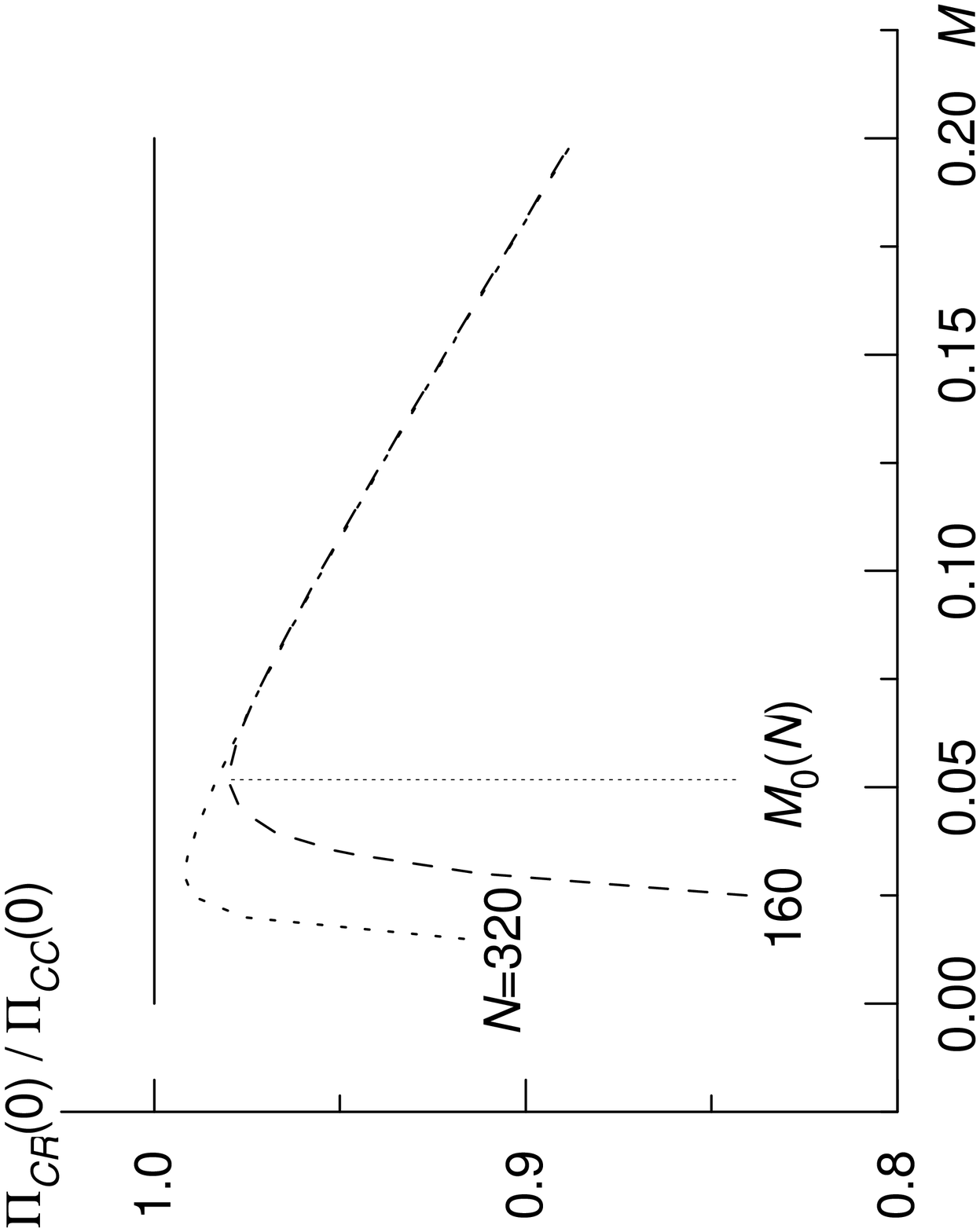}}

\vspace*{-48mm}
{\bf Fig.5.}
11112 regularized polarization operators $\Pi_{CR}(0)/\Pi_{CC}(0)$ \\
\hspace*{16mm} as functions of $M$: \\
\hspace*{16mm} 1 -- $N=160$,   \qquad  2 -- 320
\bigskip
\end{figure}

It follows from (\ref{13}), that only diagrams with more than two external
lines contribute to the arguments of the lattice and continuum determinants.
Such diagrams on the lattice differ from the corresponding continuum ones
by the terms of order ${\rm O}(1/N)$. So in the framework of perturbation
theory
\begin{equation}\label{14}
\Arg D_{CR} = \Arg D_{CW} \to \Arg D_{CC},
\qquad \mbox{ when } N \!\to\! \infty.
\end{equation}
Computer simulations of the determinant argument were performed for $N$=32,
160 and are presented at Fig.4. One sees that eq.(\ref{14}) is valid in the
interval $-0.5 \!\le\! h_1 \!\le\! 1.5$.

\begin{figure}[tb]
\epsfxsize=16cm

\vspace*{-7mm}
\hspace*{-3mm}
\rotate[r]{\epsfbox{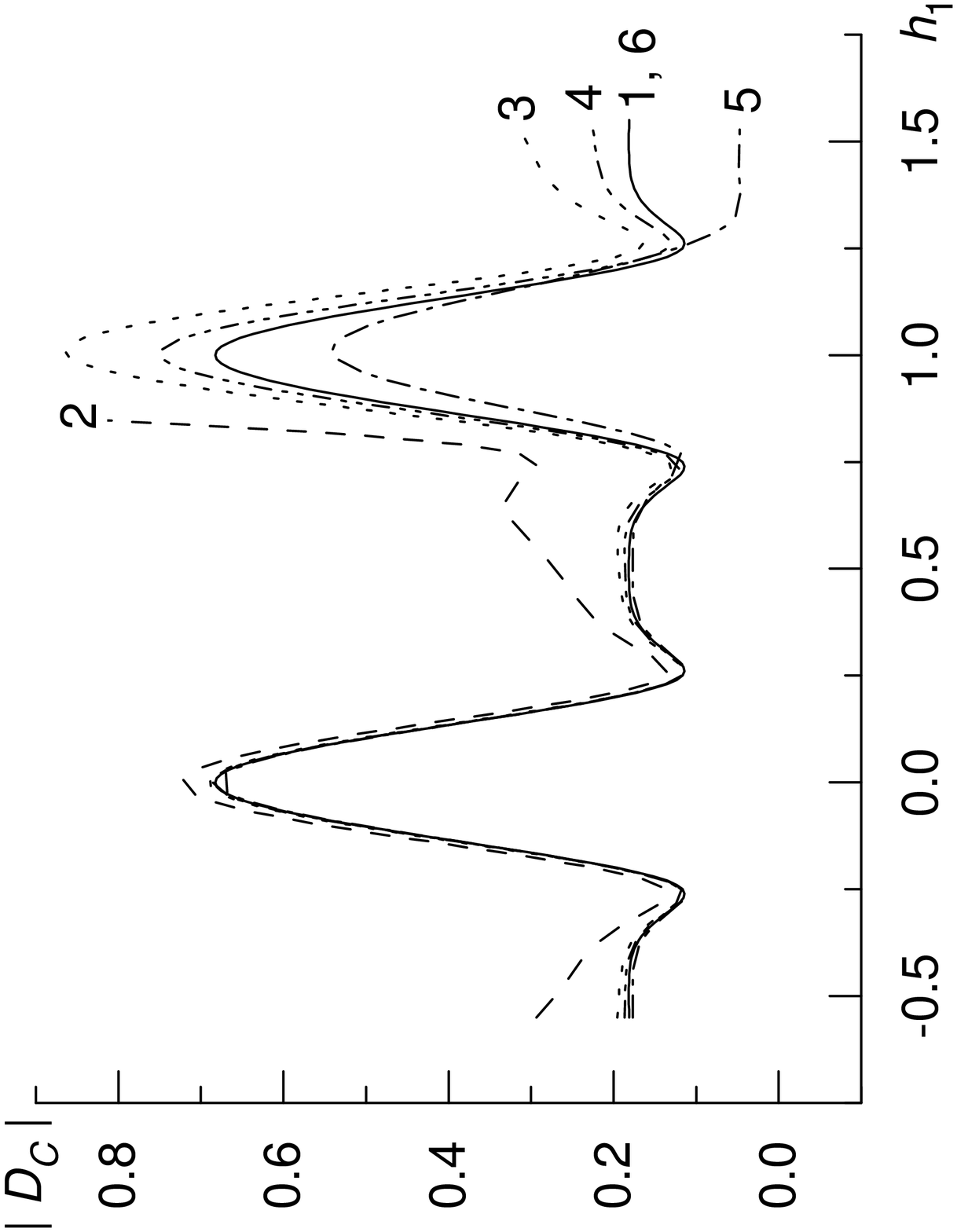}}

\vspace*{-48mm}
{\bf Fig.6.}
11112 determinant modulus $|D_C|$ as functions of $h_1$ at $h_0=0.2$: \\
\hspace*{16mm} 1 -- $|D_{CC}|$ on the torus; \qquad
2, 3, 4 -- $|D_{CR}|$ with PV field, $M=M_0$: \\
\hspace*{16mm} 2 -- $N=32$, \qquad  3 -- 160, \qquad  4 -- 320; \\
\hspace*{16mm} 5, 6 -- $|D_{CK}|$ with counterterm: \quad   5 -- $N=32$,
\quad 6 -- 160
\bigskip
\end{figure}

Since the higher order diagrams are convergent the best agreement for the
determinant modulus is achieved at the values of $M$ for which the regularized
polarization operator $\Pi_{CR}(0)$ is closest to the continuum value
$\Pi_{CC}(0)\!=\!8\pi$. Fig.5 shows that the ratio $\Pi_{CR}(0)/\Pi_{CC}(0)$
for $N \!\ge\! 160$ differs from 1 less than by 2\% if $M \!=\! M_0(N)
\!\sim\! N^{-3/4}$. This gives $M_0$=0.03 -- 0.05 at $N$=160 -- 320. It
agrees with the analytical estimates which gives
$$
\Pi_{CR}(0) = 8\pi + {\rm O}(1/MN) + {\rm O}(M^2 \ln^2 N), \qquad
\mbox{if } M \!\to\! 0 \mbox{ when } N \!\to\! \infty.
$$

The dependence of the 11112 regularized determinant modulus on $h_1$ at
$h_0$=0.2 for these values of $M$ is given at Fig.6 (curves 1 -- 4). One
sees that results obtained for the lattice Wilson action with PV
regularization agree with the continuum theory in a certain interval. When the
value $N$ grows from 32 to 320 this interval expands from $|h_1| \!\le\! 0.25$
to $-0.5 \!\le\! h_1 \!\le\! 0.9$.

As in the case of vectorial theory another way to achieve the agreement of
lattice and continuum 11112 theories is to introduce to the action (\ref{7})
a real gauge noninvariant counterterm. Then we get for the 11112 lattice
determinant
$$
D_{CK} = D_{CW} \exp\left[ k_C (h_0^2 + h_1^2)\right],
$$
where $D_{CW}$ defined above. Choosing $k_C$=6.4005 we get the curves 5, 6
shown at Fig.6. At $N$=160 one has an agreement of the theory with
counterterm and continuum one in the interval $-0.5 \!\le\! h_1 \!\le\! 1.5$.

\section {Discussion}

In this paper we showed that introducing to the fermion lattice action
additional Pauli -- Villars type regularization one can supress the gauge
symmetry breaking effects caused by the Wilson term both in perturbative and
nonperturbative regime. No gauge noninvariant counterterms are needed to get
the correct continuum result. For a finite lattice spacing there are
symmetry breaking effects of order $a$. In principle one can avoid the
symmetry breaking completely by introducing a manifestly chiral gauge
invariant action like SLAC model supplemented by PV regularization. Our
calculations show that in the vectorial model it produces a manifestly
gauge invariant result in a good agreement with the continuum expression.
However in the anomaly free chiral models there are some problems related to
the nonlocality of the SLAC model. These problems will be discussed in a
separate publication.
$$ ~ $$
{\bf Aknowlegements} \\

The authors are grateful V.Bornyakov and S.Zenkin for fruitful discussions.
This work was supported by RBRF under Grant 96-01-005511, and by INTAS Grant
INTAS-96-370.
$$ ~ $$


\begin{thebibliography}{99}
\bibitem{NiNe} H.B.Nielsen, M.Ninomiya, Nucl.Phys. {\bf B}105 (1981) 219.
\bibitem{Sh} Y.Shamir, Nucl.Phys.(Proc.Suppl.) {\bf B}47 (1996) 212.
\bibitem{FSl} S.A.Frolov, A.A.Slavnov, Nucl.Phys. {\bf B}411 (1994) 647.
\bibitem{W} K.Wilson, Phys.Rev. {\bf D}10 (1974) 2445; in "New phenomena in
subnuclear physics", ed. A.Zichichi, Plenum Press, NY, 1977.
\bibitem{AGMV} L.Alvarez-Gaume, G.Moore, C.Vafa, Comm.Math.Phys. 6 (1986) 1.
\bibitem{NNe} R.Narayanan, H.Neuberger, Phys.Lett. {\bf B}348 (1995) 549.
\bibitem{FRD} C.D.Fosco, S.Randjbar-Daemi, Phys.lett. {\bf B}354 (1995) 383.
\bibitem{Roma} A.Borelli, L.Maiani, G.Rossi, R.Sisto, M.Testa,
Nucl.Phys. {\bf B}333 (1990) 335.
\bibitem{BHS} W.Bock, J.E.Hetrick, J.Smit, Nucl.Phys. {\bf B}437 (1995) 585.
\bibitem{Sl} A.A.Slavnov, Phys.Lett. {\bf B}438 (1995) 553.
\bibitem{FrSl} S.A.Frolov, A.A.Slavnov, Phys.Lett. {\bf B}309 (1993) 344.
\end {thebibliography}
\end{document}